\documentclass[showpacs,prl,twocolumn,superscriptaddress,floatfix,amsmath]{revtex4}
\usepackage{graphicx,amsmath}
\begin{document}
\title{Excitonic condensation in a symmetric electron-hole bilayer}
\author{S. De Palo}
\altaffiliation[Present address: ]{INFM -Istituto Nazionale di Fisica della
Materia - Unit\`a di Roma}
\affiliation{INFM - Istituto Nazionale di Fisica della Materia - Unit\`a di
Trieste} 
\author{F. Rapisarda}
\altaffiliation[Present address: ]{Product and Business Development Group,
Banca IMI, Milano, Italy} 
\affiliation{Institut fur Theoretische Physik, Johannes Kepler Universitat
,Altenberger Strasse 69, A-0404 Linz, Austria}
\author{Gaetano Senatore}
\email[]{senatore@ts.infn.it} 
\affiliation{INFM - Istituto Nazionale di Fisica della Materia - Unit\`a di
Trieste} 
\affiliation{Dipartimento di Fisica Teorica, Universit\`a di Trieste,
Strada Costiera 11, 34014 Trieste, Italy}
\begin{abstract} 
Using Diffusion Monte Carlo simulations we have investigated the ground state
of a symmetric electron-hole bilayer and determined its phase diagram at
$T=0$. We find clear evidence of an excitonic condensate, whose stability
however is affected by in-layer electronic correlation. This stabilizes the
electron-hole plasma at large values of the density or inter-layer distance,
and the Wigner crystal at low density and large distance.  We have also
estimated pair correlation functions and low order density matrices, to give a
microscopic characterization of correlations, as well as to try and estimate
the condensate fraction.
\end{abstract}
\date{23 January, 2002}
\maketitle

Electron-hole systems have attracted a lot of interest over the
years~\cite{TMRice}. The Coulomb attraction existing between the two kinds of
Fermions naturally brings about pairing, and hence the possibility of a
coherent state~\cite{Keldish}. It was soon realized~\cite{Lozo} that systems of
spatially separated electrons and holes, such as a bilayer, have a number of
advantages with respect to conventional bulk samples~\cite{TMRice} in which
electrons and holes occupy the same region. Thus, while in a homogeneous
semiconductor the excitonic condensate would be an insulator~\cite{Keldish}, in
a bilayer superconductivity is in principle possible~\cite{Lozo}. The interest
in such systems has greatly increased in recent years, due to the increasing
ability to manufacture high quality semiconductor quantum well (QW)
structures, where electrons and holes are indeed confined in different
regions, between which tunneling can be made negligible~\cite{spe1}. Also, in a
number of such systems experimental evidence of an excitonic condensate has
been claimed~\cite{AlAs/GaAs,GaAS/Al$_x$Ga$1-x$As,InAs/GaSb}. However, even for
the simplest two dimensional (2D) model, i.e., a symmetric bilayer, the theory
has been concerned so far with mean-field treatments~\cite{Lozo,Zhu,Lozo2}
based on a BCS like wavefunction, with one exception~\cite{chan}.

In this Letter we report on the first extensive computer simulations of a
symmetric electron-hole bilayer (SEHB). We have undertaken this study to
assess the effect of electron correlation on the phase diagram of the SEHB and
on the existence of a condensate, which we directly characterize. To perform
simulations, we have resorted to fixed-node diffusion Monte Carlo
(FN-DMC)~\cite{dmcfn,cornell}, a method which is stable and variational and is
known to yield extremely accurate results for homogeneous electron
systems~\cite{kwon,varsano}. We should stress that our goal is to determine
the properties of the simplest 2D model, i.e., the SEHB, with unprecedented
accuracy, also to provide a benchmark against which approximate many-body
treatments may be tested. Therefore, we are not considering here effects such
as finite layer thickness, inter-layer tunneling, or band anisotropy, which
may play an important role~\cite{Lozo,Sche,Conti} in describing more realistic
QW structures.  Also, to limit the computational load, we have restricted our
study to three phases: the excitonic phase (EP), the spin unpolarized two
component plasma (2CP), and the triangular Wigner crystal (WC).

In the absence of magnetic fields, the Hamiltonian of such an ideal SEHB reads
\begin{eqnarray}
H_{eh}= -\sum_{i}\frac{\nabla^2_{i,e}}{2m_e} -
 \sum_{i}\frac{\nabla^2_{i,h}}{2m_h} +\sum_{i<
 j}\frac{e^2}{\epsilon|\mathbf{r}_{i,e}-\mathbf{r}_{j,e}|} \nonumber\\
 +\sum_{i< j}\frac{e^2}{\epsilon|\mathbf{r}_{i,h}-\mathbf{r}_{j,h}|}-
 \sum_{i,j}\frac{e^2}{\epsilon
 \sqrt{|\mathbf{r}_{i,e}-\mathbf{r}_{j,h}|^2+d^2}},
\label{Hamrs}
\end{eqnarray}
with $\epsilon$ the background dielectric constant, $d$ the inter-layer
distance, and $m_e=m_h=m^*$ the common effective mass of electrons and holes.
In the following we use $Ry^*= e^2/2\epsilon a_B^*$ as unit energy,
$a_B^*=\hbar^2 \epsilon/m^*e^2$, and $r_sa_B^*$ as unit length. As is well
known the parameter $r_s$, defined in terms of the in-layer areal density $n$
by $\pi r_s^2 {a_B^*}^2=1/n$, measures the in-layer coupling strength. In the
SEHB one is lead to define an additional parameter, measuring the importance
of the inter-layer coupling, as the ratio of the typical inter-layer and
in-layer Coulomb energies, namely $\gamma=1/d$.  At $T=0$, which is the case
considered here, the model is completely  specified by $r_s$ and $d$ or
$\gamma$.

In FN-DMC~\cite{dmcfn} one propagates a trial wavefunction $\Psi_T$ in
imaginary time, to project out the higher energy components and sample the
lowest energy state $\Psi_0$ with the nodal structure of $\Psi_T$.  This
establishes a correspondence between nodal structure, i.e., trial function,
and phase.  To study the EP we resort to a BCS-like trial function, which is
known to provide a good mean-field description ~\cite{Keldish,Comt,Zhu} both
at high and at low density. In practice we set $ \Psi_{T}^{EP} =
D_{\uparrow\uparrow} D_{\downarrow\downarrow}J$, where $D_{\sigma\sigma}$ is a
determinant of pair orbitals $ \varphi({\bf r}^e_{i,\sigma}-{\bf
r}^h_{j,\sigma})$ and $J$ is a Jastrow factor, accounting for two-body
correlations. We choose $ \varphi(r)$ as the exact numerical solution of the
mean-field problem~\cite{Zhu}, as in selected cases we found that this yields
a lower variational energy than other choices~\cite{phi}. For the normal
phases the trial function is taken as $\Psi_{T}=D^e_\downarrow D^e_\uparrow
D^h_\downarrow D^h_\uparrow J$, with $ D^a_\sigma $ a Slater determinant of
one-particle orbitals (plane waves for the 2CP and gaussians localized at the
crystal sites\cite{TWC} for the WC). The Jastrow factor
$J=\exp[-(1/2)\sum_{\mu,\nu}\sum'_{i_{\mu},i_{\nu}}
u_{\mu,\nu}(r_{i_{\mu},i_{\nu}})]$, with the Greek index denoting particle
type and spin projection, is built using RPA
pseudopotentials~\cite{rapit,J_exci}.

\begin{table}
\caption
{Energy per particle of various phases of the SEHB, in $Ry^*$, according to
FN-DMC. All results are extrapolated to the thermodynamic limit~\cite{Ext}. }
\begin{tabular}{r c l l l l}
\hline
\hline
$~~r_s$ & $d$ & $~r_E$ & $\quad$2CP & $\quad$E & $\quad$WC\\
\hline
1 &  0.0 &  1.69 & -0.833(8)  & -0.808(9)  &  \\
\hline
\hline
2 &  0.1 &  0.84 & -0.6947(5) & -0.6976(7)   &  \\
2 &  0.2 &  1.00 & -0.6116(7) & -0.6006(7)  &  \\
2 &  0.5 & 1.34 & -0.5405(5) & -0.5260(4) &  \\
\hline
\hline
5 &  0.2 & 0.57  & -0.3732(6) & -0.3822(2)  &  \\
5 &  0.5 &  0.93  & -0.3125(2) & -0.3104(1)  &  \\
5 &  1.0 &  1.40  & -0.3009(2) & -0.2987(2)  &  \\
\hline
\hline
10 &  0.5 &  0.68  & -0.1801(1)  & -0.18172(9)  &  \\
10 &  1.0 &  1.23  & -0.17153(8) & -0.17085(4)  &  \\
10& 1.5   && -0.17035(1) &       &       -0.16947(1)   \\
\hline
\hline
20 &  0.05 &  0.14   & -0.3275(2)   &  -0.3288(1) &          \\
20 &  0.5 &  0.50   & -0.10051(5) &  -0.10227(3) & -0.10157(2) \\
20 & 1.0 &  1.08   & -0.09304(1)  &  -0.09316(2)  & -0.093176(8) \\
20 &  1.3 &  1.38   & -0.09264(2)  &  -0.09261(2)    &          \\
20 &  1.5 &  1.51   & -0.09260(2) & -0.09248(2)    & -0.092533(5) \\
20 &  3.0 &     & -0.09245(2)     &                      & -0.092354(6) \\
\hline
\hline
22 & 1.0 &     &-0.08533(2)  &   &-0.085484(7) \\
22 & 2.0 &     &-0.08482(1)  &   &-0.084770(6) \\

22 & 3.0 &     &-0.08478(1)  &   &-0.084767(5) \\
\hline
\hline
30 &  0.5 &  0.45   &          & -0.07192(2) & -0.07161(2)  \\
30 &  1.0 &  1.00   &          & -0.064403(8) & -0.064483(3) \\
30 &  1.5 &  1.48    &          & -0.063833(6) & -0.063921(3) \\
30 & 3.0  &    & -0.06377(7)  &                      & -0.063827(3) \\
\hline
\hline
\end{tabular}
\label{TabE}
\end{table}

We have performed DMC simulations for systems with $N=58$ and $N=56$ particles
per layer, respectively for the 2CP and the EP and for the WC, using periodic
boundary conditions. Finite size effects have been mitigated as
usual~\cite{dc78} by performing the Ewald summation on the infinite periodic
replicas of the simulation {\it box} and using, for the 2CP, numbers of
particles  corresponding to closed shells of orbitals. We have also carried
out variational Monte Carlo (VMC) simulations, for several values of $N$
(ranging up to 114 or 120 depending on the phase), to determine the size
dependence~\cite{Ext} of the energy. Assuming that this is the same for VMC and
FN-DMC~\cite{Tana,kwon,Cortona}, we have obtained the FN-DMC energies in the
thermodynamic limit, which we report~\cite{hartree} in Table \ref{TabE} for the
three phases that we have studied.

\begin{figure}
\null\vspace{-10mm}
\includegraphics[width=67mm,angle=-90]{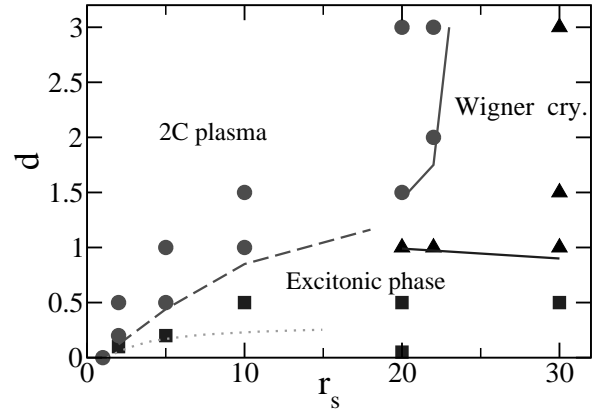}
\vspace{-5mm}
\caption{Phase diagram of the SEHB according to the FN-DMC.  Circles, squares
and triangles respectively indicate the stability of the 2CP, EP,and
WC. Dashed, gray and black lines give approximate boundaries between pair of
phases obtained from the crossing of the energies at given $r_s$ with varying
$d$. The dotted curve reports an estimate of the 2CP-EP boundary from an
approximate theory~\cite{Lerw}.}
\label{Fig1}
\end{figure}

Use of the energies of Table \ref{TabE} yields the phase diagram shown in
Fig. \ref{Fig1}. It is evident that correlation has here qualitative effects.
Whereas the mean-field predicts stability of the EP with respect to the 2CP
everywhere, at high enough density and/or at large enough distance FN-DMC
predicts the stability of either the 2CP or the WC. Naively, one would expect
that when the inter-layer coupling $\gamma=1/d \ll 1$ each layer should behave
as an isolated 2D electron gas (2DEG)~\cite{SJ}. This is the case with the
symmetric electron bilayer~\cite{rapit,Rapi,Con2}, for which the phase diagram
of the 2DEG~\cite{Rapi} is quickly recovered as $d$ exceeds 1, and it appears
to be the case also for the SEHB, when one keeps in mind that only the
unpolarized 2CP is considered here.  The phase diagram turns out to be fairly
robust. Thus, use of the finite-$N$ FN-DMC energies leaves it essentially
unchanged, while use of the VMC energies only brings about a minor change in
the boundary between the WC and the 2CP, which moves at about $r_s=20$ for
$d\ge 1.5$.

In Table \ref{TabE} we have also reported the excitonic radius, defined by
$r_E^2=\langle \varphi |r^2|\varphi\rangle/\langle \varphi |\varphi\rangle$.
It is evident, from an inspection of the Table, that a correlation exists
between the stability of the EP and $r_E$ being smaller than 1, i.e., smaller
than the characteristic in-layer length scale.  This points to the importance
of in-layer correlations, which are neglected in mean-field~\cite{Lozo,Zhu}.
Also, it turns out that in our simulations the {\it exciton} is always much
smaller than the side of the simulation {\it box}, $\sqrt{N\pi}\simeq
13$. Thus, our description of the EP should not be affected by the finite
spacing of the energy levels~\cite{chan}.

\begin{figure}
\includegraphics[angle=-90,width=9.5cm]{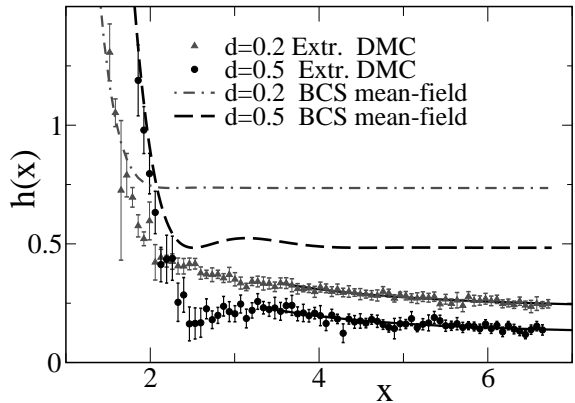}
\vspace{-9mm}
\caption{Projected two-body density matrix of the SEHB at $r_s=5$, according
to FN-DMC and BCS mean-field, for $N=58$. The full curves are fits to the
tails of the simulation data (see text and Table \ref{TabC} for details).}
\label{f:Fig2}
\end{figure}

\begin{table}
\caption
{Condensate fraction $\alpha$ of the SEHB according to FN-DMC (extrapolated
estimate) and VMC, for a system with N=58. Here $\tilde{d}=d\cdot r_s$ is the
interlayer distance in units of $a_B^*$. Also reported are the BCS mean-field
prediction and, for FN-DMC, the reduced $\chi^2$ of the fit yielding the
condensate fraction (see text).}
\begin{tabular}{r  c c c c c }
\hline
\hline
$r_s$  &$\tilde{d}$&  $\alpha$(BCS)  & $\alpha$(VMC) & $\alpha$(FN-DMC)  & $\quad\chi^2$\\
\hline
2 &    0.2  & 0.55 & 0.187(4)    & 0.284(9)   & 0.59  \\
5 &   1.0  & 0.74 & 0.151(2)   & 0.215(4) &0.22     \\
5 &   2.5 & 0.48 & 0.095(3)   & 0.108(5) &0.34   \\
20&    1.0  & 0.98 & 0.027(1)    & 0.020(2) & 0.47  \\
\hline
\hline
\end{tabular}
\label{TabC}
\end{table}

A peculiar property of superconductors is the off-diagonal long-range order
(ODLRO) exhibited by the reduced density matrices in the coordinate space
representation~\cite{condensate}. In a Fermion system ODLRO shows up in the
two-body density matrix~\cite{condensate}, with the appearance of an eigenvalue
which scales with the number of particles. In a translational invariant system,
such as the SEHB in the excitonic phase, ODLRO implies for the two-body density
matrix, to leading order in $N$, the following asymptotic behavior:
\begin{eqnarray}
\label{eq-cond}
\rho_2({\bf x}'_e, {\bf x}'_h;
  {\bf x}_e, {\bf x}_h)= \alpha N f^{*}(
|{\bf x}'_e-{\bf x}'_h|)f(|{\bf x}_e-{\bf x}_h|),\\
|{\bf x}_e-{\bf x}_h|, |{\bf x}'_e-{\bf x}'_h|
\alt \xi, \quad |{\bf x}_e-{\bf x}'_e|  \rightarrow \infty,\nonumber
\end{eqnarray}
where $\alpha \le 1$ is the condensate fraction and $\xi$ is the range of the
normalized pair amplitude $f(|{\bf x}_1 -{\bf x}_2|)$. In order to estimate
the condensate fraction it is convenient to resort to the projected two-body
density matrix (P2BDM)
\begin{eqnarray}
h(\mathrm{x})=
\frac{1}{N}\int d{\bf x}_ed{\bf x}_h \rho_2({\bf x}_e+{\bf x}, {\bf
 x}_h+{\bf x}; {\bf x}_e, {\bf x}_h),
\label{eq-projected}
\end{eqnarray}
which tends to $\alpha$ in the large $\mathrm{x}$ limit, as is immediately
found combining Eqs. (\ref{eq-cond}) and ({\ref{eq-projected}).

A simple estimator of $h(\mathrm{x})$ is given by 
\begin{equation}h(\mathrm{x}) =  \frac{N}{M_c}\sum_{i=1}^{M_c}
\frac{\Psi_T({\bf R}_i')}
{\Psi_{T}({\bf R}_i)},
\label{eq-estimator}
\end{equation}
with $M_c$ the number of particle configurations used and ${\bf R}'$ obtained
from the configuration ${\bf R}$ by rigidly translating an electron-hole
pair~\cite{size-cor} by ${\bf x}$. In practice, for each ${\bf R}$ we generate
a few translations ${\bf x}$ uniformly distributed in the simulation {\it box}
and, for better statistics, we also average Eq. (\ref{eq-estimator}) over all
electron-hole pairs. As Eq. (\ref{eq-estimator}) only yields a mixed estimate
in DMC, we have also performed VMC calculations to get extrapolated
estimates~\cite{Ceperley79} according to
$h_{Extr}(\mathrm{x})=2h_{DMC}(\mathrm{x})-h_{VMC}(\mathrm{x})$.

\begin{figure}
\includegraphics[angle=-90,width=8.5cm]{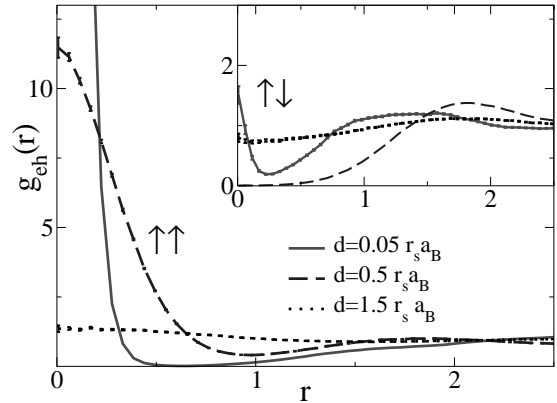}
\vspace{-1mm}
\caption{ Spin resolved electron-hole pair correlation function of the SEHB at
$r_s=20$, according to FN-DMC.}
\label{f:Fig3}
\end{figure}

An illustration of our results for $h(\mathrm{x})$ is given in
Fig. \ref{f:Fig2}. Indeed, $h(\mathrm{x})$ appears to saturate at large
$\mathrm{x}$.  In the absence of exact information on its asymptotic form, we
have fitted its tail to $\alpha+A/\mathrm{x}^2$, $\mathrm{x}\ge 5$ and
estimated the condensate fractions reported in Table \ref{TabC} for the cases
studied.  It is evident that in-layer correlation, which is absent in mean
field, causes a substantial reduction of the condensate fraction---a reduction
which becomes more pronounced with increasing the in-layer coupling $r_s$, at
given interlayer distance, and to less extent with increasing the interlayer
distance, at given $r_s$. This effect is particularly strong at large
coupling, yielding the essential suppression of the condensate at $r_s=20$ and
$d=0.05$, with $\alpha\simeq 0.02$ whereas in mean field $\alpha\simeq 1$.

Additional insight into the nature of the phases of the SEHB is provided by
the pair correlation functions, whose features we briefly summarize here.  As
the two layers are brought together from infinity inter-layer interaction
produces a substantial buildup in inter-layer correlations and the screening
and weakening of intra-layer correlations, as for the electron
bilayer~\cite{rapit,Rapi,Con2}. However, while $g_{eh}(r)$ and
$g_{eh}^{\uparrow\uparrow}(r)$ monotonously increase with decreasing $d$, near
the origin, $g_{eh}^{\uparrow\downarrow}(r)$, first develops a {\it
correlation hole} and then by further diminishing $d$ develops a peak.  This
behavior, which becomes particularly marked at large coupling as is evident
from Fig. \ref{f:Fig3}, might be interpreted as a tendency toward biexciton
formation for large $r_s$ and small $d$. In fact,
$g_{eh}^{\uparrow\downarrow}(r)$ gives correlations between unpaired
($\uparrow\downarrow$) electrons and holes and, indirectly, between $\uparrow$
and $\downarrow$ excitons.

To summarize, by performing extensive quantum simulations we have shown that
correlation has important qualitative effects in determining the phase diagram
and the excitonic condensate of the SEHB, with respect to mean-field. Here, we
have chosen not to consider spin polarized phases nor inhomogeneities of spin
or charge, such as in density waves and in liquid-vapor coexistence. We shall
explore some of these interesting phenomena in future investigations.

We acknowledge useful discussions with many colleagues and among the others
with D. Neilson, C. Castellani, R.K. Moudgil.  SDP and GS respectively
acknowledge the holding of INFM fellowships while carrying out this work and
support by MURST through COFIN99.


\begin{thebibliography}{99}
\bibitem{TMRice}T.M.Rice, Solid State Phys. {\bf 32},
1 (1977); J.C.Hensel, T.G Phillips, and G.A. Thomas, {\sl ibid.} {\bf 32}, 88
(1977).
\bibitem{Keldish}L.V. Keldish and Y.V. Kopaev, Fiz. Tverd. Tela {\bf 6} 2791
(1964) [Sov. Phys. Solid State {\bf 6}, 2219 (1965)]; A.N.Kozlov and
L.A. Maximov, Zh. Eksp. Teor. Fiz. {\bf 48}, 1184 (1965)[Sov. Phys. JEPT {\bf
21}, 790 (1965)]; L.V. Keldish and A.N.Kozlov, Zh. Eksp. Teor. Fiz. {\bf 54},
1978 (1968)[Sov. Phys. JEPT  {\bf 27}, 521 (1968)].
\bibitem{Lozo}
 Yu. E. Lozovik and V. I. Yudson, Pis'ma Zh. Eksp. Teor.
Fiz. {\bf 22}, 556 (1975)[JETP Lett. {\bf 22}, 274 (1975)]; Solid State
Commun. {\bf 19}, 391 (1976); Zh. Eksp. Teor. Fiz. {\bf 71}, 738 (1976)
[Sov. Phys. JETP {\bf 44}, 389 (1976)].
\bibitem{spe1} 
U.Sivan, P.M. Solomon and H. Shtrikman, Phys. Rev. Lett {\bf 68}, 1196 (1992);
B. Kane {\it et al.}, Appl. Phys. Lett. {\bf 65}, 3266 (1994).
\bibitem{AlAs/GaAs}
L.V. Butov  {\it et al.} Phys. Rev. Lett. {\bf 73}, 304 (1994).
\bibitem{GaAS/Al$_x$Ga$1-x$As}
V.B. Timofeev {\it et al.} Phys. Rev. {\bf B 61}, 8420 (2000).
\bibitem{InAs/GaSb}
J. P Cheng {\it et al.}, Phys. Rev. Lett. {\bf 74}, 450 (1995)
J. Kono {\it et al.}, Phys. Rev. {\bf B 55}, 1617 (1997);
T.P. Marlow  {\it et al.}, Phys. Rev. Lett. {\bf 82}, 2362 (1999).
\bibitem{Lozo2} Y.E. Lozovik and O.L. Berman, Physica Scrip. {\bf 55}, 491
(1997).
\bibitem{Zhu} Xuejun Zhu, P.B. Littlewood, S. Hybersten and T. M. Rice,
 Phys. Rev. Lett {\bf 74}, 1633 (1995),P.B. Littlewood and Xuejun Zhu,
Phys. Scripta  {\bf T 68}, 56 (1996).  
\bibitem{chan} Antony Chan, PhD Thesis, Indiana University (1996).
\bibitem{dmcfn} P. J. Reynolds, D. M. Ceperley, B. J. Alder and W.A. Lester,
J. Chem. Phys. {\bf 77}, 5593 (1982); see also C. J. Umrigar,
M.P. Nightngale and K.J Runge, J. Chem. Phys. {\bf 99}, 2865 (1993). 
\bibitem{cornell}For a recent survey of QMC methods see, e.g.,
{\it Quantum Monte Carlo Methods in Physics and Chemistry},
ed. M.P. Nightingale and C.J. Umrigar (Kluwer, Dordrecht, 1999).
\bibitem{kwon} Y. Kwon, D.M Ceperley, and R.M.  Martin, Phys. Rev. B {\bf 53},
7376 (1996); {\sl ibid.} {\bf 58}, 6800 (1998).
\bibitem{varsano}D.Varsano, S.Moroni, and G.Senatore, Europhys. Lett. {\bf
53}, 348 (2001).
\bibitem{Sche} S.I. Shevchenko, Phys. Rev. Lett. {\bf 72}, 3242 (1994).
\bibitem{Conti} S.Conti, G. Vignale and A.H. MacDonald, Phys. Rev. 
{\bf B 57}, 6846 (1998).
\bibitem{Comt} C. Comte and P.Nozieres, J. Phys. {\bf 43}, 1069 (1982); 
P.Nozieres and C. Comte, {\it ibid.} {\bf 43}, 1083 (1982).
\bibitem{phi} We tried a variational gaussian and the form of
Ref.~\onlinecite{chan}.
\bibitem{TWC} The electron and hole triangular lattices  are chosen in
 the energetically favored AA stacking.  
\bibitem{rapit} F. Rapisarda, PhD Thesis, Universit\`a di Trieste (1995).
\bibitem{J_exci}
For the EP $J$ is constructed using the 
2CP  pseudopotentials.  Depending on  coupling the
pseudopotentials between paired particles or even all  interlayer
pseudopotentials may be dropped, to optimize the variational
energy. 
\bibitem{dc78}
D.Ceperley, Phys. Rev. B {\bf 18}, 3126 (1978). 
\bibitem{Ext}The functional forms of Ref.~\onlinecite{Tana}, which have proven
satisfactory also for the electron bilayer~\cite{rapit}, have been used for the
normal phases and with minor adaptions (like
$E_N=E_\infty+b_2(r_s)/N^\alpha(r_s)$ ) also for the EP.
\bibitem{Tana} B. Tanatar and D.M. Ceperley, Phys. Rev. {\bf B 39}, 5005
(1989).
\bibitem{Cortona}
G. Senatore, S. Moroni, and D. Varsano, Solid. St. Comm. {\bf 119} 333 (2001).
\bibitem{hartree} Our energies do not include the phase independent Hartree
term $2d/r_s$, accounting for the electric field present 
if the layers  are not separately neutralized.
\bibitem{SJ} We only consider 2DEG studies using Slater-Jastrow trial
functions---the same chosen here for the normal phases.
\bibitem{Rapi} F. Rapisarda and G. Senatore, Aust. J. Phys. {\bf 49}, 161
(1996).
\bibitem{Con2} G. Senatore, F.Rapisarda and S. Conti,
Int. Jour. Mod. Phys. {\bf B 13 },479 (1999). 
\bibitem{Lerw} Lerwen  Liu, L. Swierkowsky and D.Neilson, 
Physica {\bf B 249}, 594 (1998).
\bibitem{condensate} C.N. Yang, Rev. Mod. Phys. {\bf 34}, 694  (1962).  
\bibitem{size-cor}To reduce size effects only the pair in the simulation {\it
box} is displaced, while its periodic copies are held fixed. We have indeed
found that this procedure, which is an obvious extension of the one introduced
in Ref. \onlinecite{Magr} for the one-body density matrix, substantially
reduces size effects in VMC simulations with $N=42,58,114$.
\bibitem{Magr} W.R Magro and D.M. Ceperley Phys. Rev. Lett. {\bf 73},
826 (1994).
\bibitem{Ceperley79} D.M.Ceperley and M.H.Kalos, in {\it Monte Carlo Methods
in Statistical Physics}, K.Binder, ed. (Springer, N.Y., 1979).
\end{thebibliography}
\end{document}